\documentclass[preprint,groupedaddress,showpacs]{revtex4}%

\usepackage{graphicx}
\usepackage{amsmath}
\usepackage{epsfig}
\usepackage{color}
\usepackage{bm}
\usepackage[sort&compress]{natbib} 

\graphicspath{{Pics/}}

\begin{document}

\title{Electromagnetic field generation in the downstream of electrostatic
shocks due to electron trapping
}

\author{A. Stockem\(^{1,2}\)}
\email[Electronic address: anne@tp4.rub.de]{}
\author{T. Grismayer\(^{2}\)}
\author{R. A. Fonseca\(^{2,3}\)}
\author{L. O. Silva\(^2\)}
\affiliation{\(^1\)Institut f\"ur Theoretische Physik, Lehrstuhl IV: Weltraum- und Astrophysik, Ruhr-Universit\"at Bochum, D-44780 Bochum, Germany\\
\(^2\)GoLP/Instituto de Plasmas e Fus\~ao Nuclear - Laborat\'orio
Associado, Instituto Superior T\'ecnico, Universidade de Lisboa, Lisboa,
Portugal\\
\(^3\)ISCTE Instituto Universit\'ario Lisboa, Portugal}

\date{\today}

\begin{abstract}
A new magnetic field generation mechanism in electrostatic shocks is found,
which can produce fields with magnetic energy density as high as 0.01 of the
kinetic energy density of the flows on time scales \(\sim 10^4\, \omega_{pe}^{-1}\). Electron
trapping during the shock formation process creates a strong temperature anisotropy in the
distribution function, giving rise to the pure Weibel instability. The generated
magnetic field is well-confined to the downstream region of the electrostatic
shock. The shock formation process is not modified and the features of the shock front responsible for ion acceleration, which are currently probed in laser-plasma laboratory
experiments, are maintained. However, such a strong magnetic field determines the
particle trajectories downstream and has the potential to modify the signatures of the
collisionless shock.
\end{abstract}



\maketitle

\clearpage

Collisionless shocks have been studied for many decades, mainly in the context
of space- and astrophysics \cite{SK91,S08,MF09,FF12}. Recently, shock
acceleration raised significant interest in the quest for a laser-based ion
acceleration scheme due to an experimentally demonstrated high beam quality
\cite{B08,FS12,R08,HT12}. Interpenetrating plasma slabs of hot electrons and cold ions are acting to set up the electrostatic fields via longitudinal plasma instabilities. The lighter electrons leaving the denser regions are held back by the electric fields, which pull the ions. Particles are trapped in the
associated electrostatic potential, which steepens and eventually reaches a
quasi-steady state collisionless electrostatic shock. Most of the theoretical work dates back to the 70's \cite{MJ69,B69,FS70,TK71,S72} relying on the pseudo Sagdeev potential \cite{S66} and progress has been mainly triggered by kinetic simulations \cite{SM06,D92,SM04,DS10}.

The short formation time scales and the one-dimensionality of
the problem make it easily accessible with theory and computer simulations. However, long time shock evolution were
often one-dimensional or electrostatic codes were used, and the role of
electromagnetic modes was mostly neglected. More advanced multidimensional simulations have
shown the importance of electromagnetic modes also in this context, due to
transverse modes which are excited on the ion time scale \cite{KT10,SD11}. We
show that in the case of very high electron temperatures associated with the formation of electrostatic shocks \cite{SF13a}, electromagnetic modes become important on electron time scales, creating strong magnetic fields in the downstream of the shock.

With the increase in laser energy and intensity, the possibility to drive electrostatic shocks has become important for laboratory experiments of electrostatic shocks. Recent laser-driven shock
experiments showed the appearance of an electromagnetic field structure
\cite{KR12,FF13,HF13}, which was attributed to the ion-filamentation instability
\cite{PR12} that evolves on time scales of ten thousands of the inverse electron plasma frequency,
\(\omega_{pe}^{-1}\). As a main outcome of this paper, we show that these
structures can already be seeded and produced on tens of \(\omega_{pe}^{-1}\)
and remain in a quasi-steady state over thousands of \(\omega_{pe}^{-1}\). In fact, on electron time scales, the magnetic field is driven by the pure Weibel instability \cite{F59,W59,B09} due
to a strong temperature anisotropy \cite{TM10} which is caused by electron
trapping in the downstream region of electrostatic shocks.

We will start by considering shock formation in a system of symmetric charge and current neutral counterstreaming beams,
each consisting of a population of hot electrons and cold ions. The fluid velocities are chosen low enough and the electron temperature
sufficiently high, so that the electron filamentation instability associated
with the counterflows evolves only on time scales orders of magnitude larger
than the shock formation time scale, according to the analysis in ref.\ \cite{SF13a}. In this case, and for non-relativistic flow
velocities, the shock formation process is dominated by electrostatic modes.

The initial stage of the shock formation process is studied in particle-in-cell
simulations with the fully relativistic code OSIRIS \cite{F02,F08}. We use a 3D
simulation box with the length in each direction being \(L_x = L_y = L_z = 60 \,
c / \omega_{pe}\), periodic boundaries in the transverse directions and a spatial
resolution \(\Delta x = 0.1 \, c/ \omega_{pe}\) in all three directions. The temporal resolution is
\(\Delta t \omega_{pe} = 0.057 \) and a cubic interpolation scheme was used with
4 particles per cell and per species. Test runs with a higher number of
particles per cell were also performed, yielding similar results.
The full shock formation for a fluid of hot electrons and cold ions with proper velocities \(u_0  = \beta_0 \gamma_0 = \pm 0.015 \), where \(\beta_0 = v_0 / c\) and upstream Lorentz factor \(\gamma_0 = (1-v_0/c)^{-1/2}\) happens on time scales \(t \gtrsim 10 \, \omega_{pi}^{-1}\), which we followed in 3D for a reduced mass ratio \(m_p / m_e = 100\). Long-term simulations with a realistic proton to electron mass ratio \(m_p / m_e =1836\) were performed in two spatial dimensions up to \(t \approx 10^4\, \omega_{pe}^{-1}\) with \(L_x = 10^3 \,
c / \omega_{pe}\), \(L_y = 450 \,
c / \omega_{pe}\), \(\Delta x = \Delta y = 0.1 \, c/ \omega_{pe}\) and 
\(\Delta t \omega_{pe} = 0.07 \). In the 2D setup the 
thermal parameter is \(\Delta \gamma = k_B T_e / m_e c^2 = 20\), whereas it could be reduced to \(\Delta \gamma = 0.015\) in the 3D case due to the lower mass ratio, guaranteeing the electrostatic character of the shock \cite{SF13a}. 

\begin{figure}
\setlength{\unitlength}{1cm}
\begin{picture}(8,8.5)
\put(0.35,2.7){
\includegraphics[width=7cm]{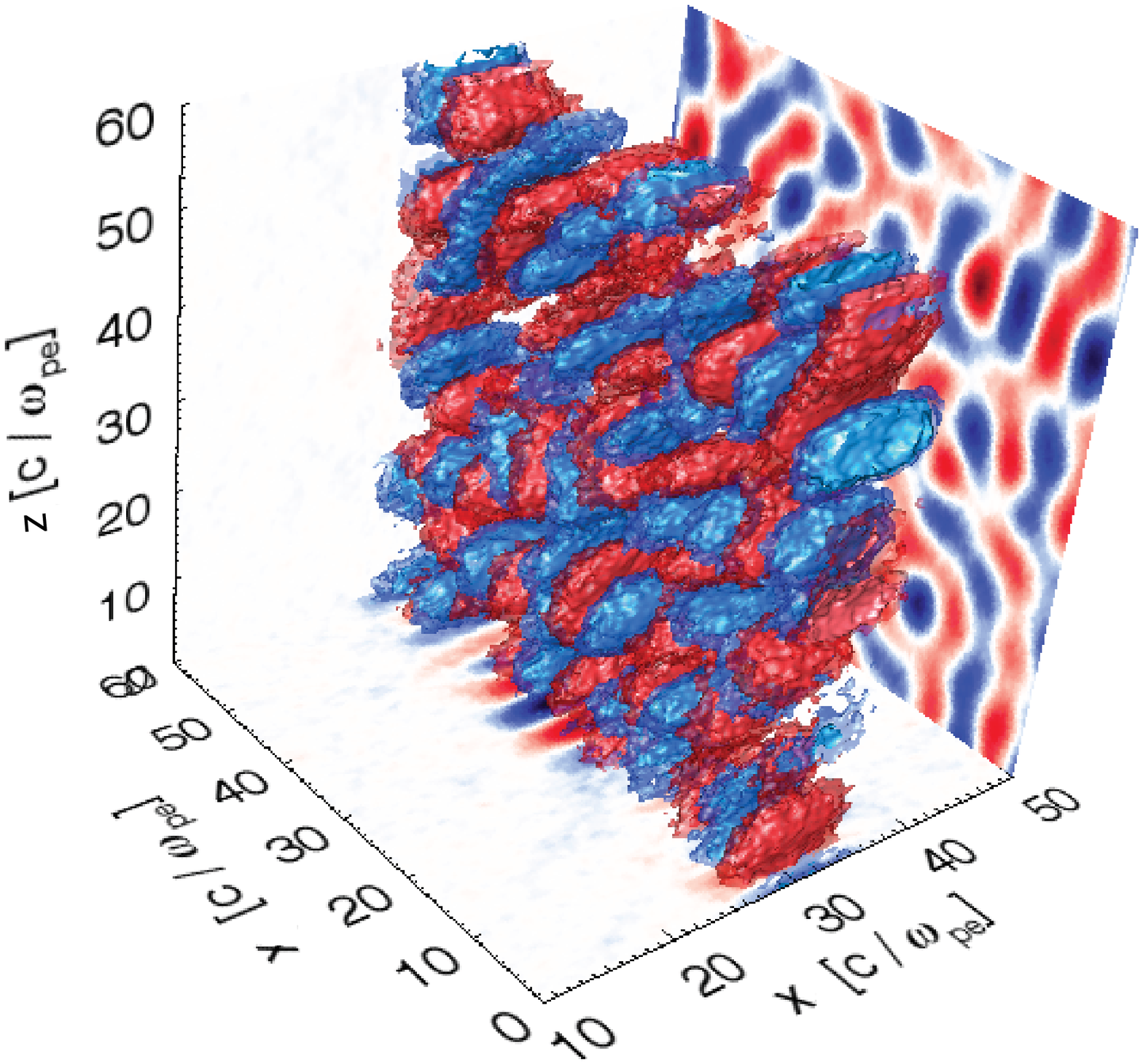}}
\put(2.,8){(a)}
\put(0,0){
\includegraphics[width=8cm]{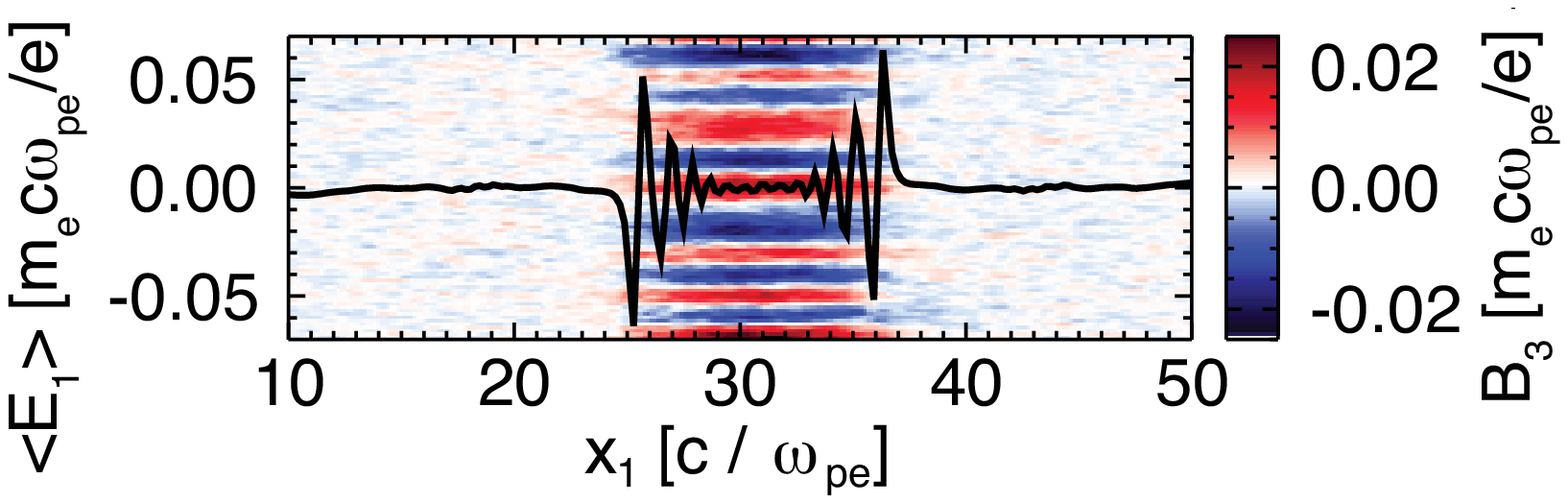}}
\put(2.2,2.1){(b)}
\end{picture}
\vspace{-12pt}
\caption{Shock formation in the 3D simulation for mass ratio \(m_p/m_e = 100\), \(u_0 = \pm 0.015\) and \(\Delta \gamma=0.015\): (a) Perpendicular electromagnetic field, (b) box-averaged electrostatic field \(<E_1>\) (black) and 2D-slice of magnetic field in (a) at \(z = 30 \, c/\omega_{pe}\) showing the extension of the filaments. Time is \(t \omega_{pe} = 460\).}\label{fig1}
\end{figure}

In this configuration similar to the configuration employed in recent experiments \cite{R08,HT12}, two symmetric shocks moving in opposite directions (along
\(x\)) are launched from the contact discontinuity at the centre of the
simulation box, where the two plasma shells initially come in contact. The
region between the two shocks defines the downstream of the two nonlinear
structures. Early in the shock formation process, we observe the generation of a
magnetic field in the downstream region between the two shock fronts. This is
illustrated in Fig.\ \ref{fig1}, where the field structure is presented after
the electrostatic shocks have reached a quasi-steady state. A strong
longitudinal electric field has already formed at the shock front. At the same
time, a strong perpendicular magnetic field has been generated, which is
well-confined to the downstream region of the shock. Unlike Weibel mediated
shocks \cite{FF12}, the magnetic field in the shock front and in the upstream
region is very small. The filamentary field structure in the downstream region
indicates the Weibel instability as the driving mechanism, reinforced by the time scales of the process (\(\sim \) tens of \( \omega_{pe}^{-1}\)) and the transverse length scale of the filaments early in time (\(\sim \) a few \( c/\omega_{pe}\)).

Several 2D simulations with proper velocities \(u_0 = 0.005- 0.1\) and \(\Delta
\gamma = 0.01-20\), corresponding to electron thermal energies \(k_B T_e\) in the range \(\sim \) 5 keV to 10 MeV, were performed in order to study the magnetic field formation
process in electrostatic shocks in more detail. This parameter range covers astrophysical conditions, e.\,g.\ with estimated quasar temperatures of \(\sim10^7 \) K \cite{Hall}, or laser-plasma interactions, where hot electrons can easily be generated with \(T_{e,hot}\) up to several MeV.

\begin{figure}[h]
\begin{center}
\includegraphics[width=7cm]{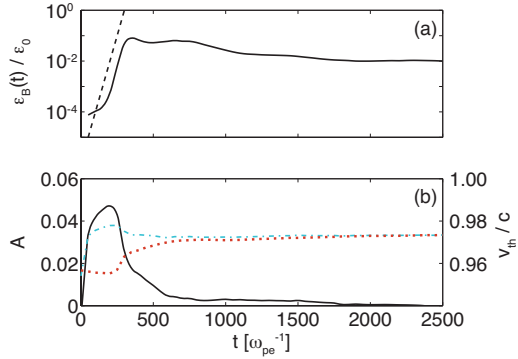}
\end{center}
\vspace{-24pt}
\caption{Temporal evolution of (a) normalised magnetic energy density and (b) thermal velocities
\(v_{th,\parallel}\) (dash-dotted), \(v_{th,\perp}\) (dotted) and anisotropy \(A\)
(solid), in a 2D simulation with \(m_p/m_e= 1836\), \(u_0 = \pm 0.1\) and \(\Delta \gamma=10\) measured over \( \Delta x_1 = 0.7 \, c / \omega_{pe}\) at the center of
the simulation box. The black dashed line in (a) is \( \exp( 2 \sigma_m t)\).  }\label{fig2}
\end{figure}

The 2D simulations reproduce the same magnetic field generation mechanism with the field
confined to the downstream region of the shock. The magnetic field energy
averaged over \(x_2\) in the centre of the shock downstream region is represented in
figure \ref{fig2}a for \(u_0 = 0.1\) and \( \Delta \gamma = 10\). After a linear
increase, at \(t \approx 300 \, \omega_{pe}^{-1}\) the field growth saturates
and a quasisteady value \(\epsilon_B \approx 0.01 \epsilon_0 \approx 0.002 (\epsilon_0 + \epsilon_{th})\) is reached, where \(\epsilon_0 = n_0 m_p (\gamma_0 -1) c^2\) represents the kinetic energy density of the ions, \(\epsilon_{th} \approx 3 \Delta \gamma / 2\) is the thermal energy density of the electrons, and \(\epsilon_B\) is the magnetic field energy density. This field structure can then seed the filamentation on the longer ion time scale, and thus sustain a high level of
\(\epsilon_B\) covering the full downstream region for times at least as long as
\(t \approx 10^4\, \omega_{pe}^{-1}\).

We now analyse the different instabilities that can arise in initially unmagnetised counter-streaming electron-ion flows. For our range of parameters, the electron current
filamentation instability is suppressed since the flows are hot \cite{Su87,SF02}. Moreover, the
cold ion-ion filamentation instability, which has been considered in connection
with recent experiments, has a maximum theoretical growth rate  \(\sigma_{i} =
\sqrt{\frac{2}{\gamma_0}} \beta_0 \omega_{pi} = 3.3 \times 10^{-3} \,
\omega_{pe}\), and a saturation field \(B \simeq m \gamma_0^2 \sigma_i^2/ q k_i
u_{x0}\) with \(\gamma_0 \) the Lorentz factor of the counterpropagating flows,
proper velocity in \(x\) direction \(u_{x0} = \beta_{x0} \gamma_0\) and wave
number \(k_i\) at the maximum growth rate \(\sigma_i \) \cite{SF02}, yielding a
saturated magnetic field of only \(B = 1.5 \times 10^{-3} \, m_e c \omega_{pe} /
e\) clearly below the field values, up to \(B \approx 2 \, m_e c \omega_{pe} /
e\), observed in the simulations. It is then clear that only an instability associated with the shock formation process can lead to magnetic field generation on the relevant time scales.

We attribute the magnetic field growth and saturation level to the
temperature anisotropy that is generated during the electrostatic shock formation
process. This is illustrated by Figure \ref{fig2}b showing the parallel and perpendicular thermal
velocities of the downstream electron distribution, \(v_{th,\parallel}\) and
\(v_{th,\perp}\) respectively with \(v_{th,\alpha} = \sqrt{k_B T_\alpha / m_e}\), together with the anisotropy parameter \(A =
(v_{th,\parallel}/ v_{th,\perp})^2-1\). In fact, the nonlinear evolution of the longitudinal modes associated with shock formation increase the electron temperature in the shock propagation direction
due to wave breaking and electron trapping, while the transverse profile of the distribution
function stays almost unchanged. At the time when the anisotropy reaches its
maximum \(A = 0.048\), the magnetic field grows exponentially at its maximum
growth rate, according to the theory for the Weibel instability \cite{W59}. Unlike previous
works, which have addressed current filamentation scenarios (with free energy
for the instability associated with non zero fluid velocities of the flows), in
this region the electron fluid velocity is zero, and the magnetic field originates only
from the temperature anisotropy associated with preferential heating along the shock formation direction \(x\) and the distortion of the distribution function due to electron trapping in the shock downstream.

We now quantify the main features of the instability driven by the electrons in the downstream of the electrostatic shock. In our model we assume only the initial conditions of the flows and that an electrostatic shock is formed \cite{SM06,SF13}. The growth rate can be calculated theoretically
from first principles. Starting from the Sagdeev description for electrostatic
shocks, the electron distribution function along the entire shock structure is calculated by
considering a free, streaming population of electrons  \( f_{e\pm} = n_0 \exp \{-[\gamma_0 (\gamma-\varphi)-1 \pm u_0 \sqrt{(\sqrt{1+u_x^2}-\varphi)^2-1}]/ \Delta \gamma\}\), as well as a trapped population in the electrostatic potential \(\varphi \), represented by a plateau in phase space for \(|u_x| \leq \sqrt{(1+\varphi)^2-1}\), \(f_{e,t} = n_0 \exp\{ -[ \gamma_0 \gamma_\perp -1] \} / \Delta \gamma\}\) with normalisation factor \(n_0\) \cite{SF13}. This distribution function is then used to evaluate the dispersion relation for electromagnetic waves \(k^2 c^2 - \omega^2 - \omega_{pe}^2 (U_e + V_e ) = 0\) with
\begin{equation}
 U_e = \int_{-\infty}^\infty d^3 u \frac{u_x}{\gamma} \frac{\partial f}{\partial u_x} \textrm{, } \quad V_e = \int_{-\infty}^\infty d^3 u \frac{u_x^2}{\gamma (\gamma \omega/kc - u_x)} \frac{\partial f}{\partial u_z}.
\end{equation}
In the non-relativistic limit
we obtain \cite{SF13a}
\begin{equation}\label{eq1}
	k^2c^2+ \sigma^2+ \omega_{pe}^2 \left[ 1- V(\phi ) \left[  1+
\frac{\imath \sigma}{\sqrt{2} v_{th}kc}\textrm Z \left(  \frac{\imath
\sigma}{\sqrt{2} v_{th}kc} \right) \right] \right] =0
\end{equation}
where \(\sigma\) is the imaginary part of the wave frequency, \(k\) is wave number along the perpendicular direction, the plasma dispersion function is \(Z\) \cite{FC61}, and \(\phi = e \varphi / m_e c^2\) is the normalised electrostatic potential with
\(V(\phi )= n_0  \left \{ e^{\sqrt{\phi}/v_{th}} \textrm{erfc}
[\sqrt{\phi}/v_{th}] + 2 \sqrt{\phi / \pi v_{th}^2} + \frac{4}{3} \sqrt{ \phi^3
/ \pi } v_{th}^{-3}\exp\left[-v_0^2 / 2v_{th}^2\right]  \right\}\) the Sagdeev potential and \( n_0 =
\left[e^{\sqrt{\phi}/v_{th}} \textrm{erfc} [\sqrt{\phi}/v_{th}]  + 2  \sqrt{\phi
/ \pi v_{th}^2}\exp\left[-v_0^2 / 2v_{th}^2\right]  \right]^{-1}\) the electron density along the shock front. At the time
when the magnetic field starts to grow, the electrostatic potential in the
downstream region is approximately of the order of the initial ion kinetic
energy \cite{fnote}, which leads to a maximum growth rate
\(\sigma_m = 0.053 \, \omega_{pe}\) from equation (\ref{eq1}) and matches well
the simulation result in figure \ref{fig2}a. The dominant wave number \(k_m =
0.14 \, \omega_{pe} / c\) corresponds to the wave length \(\lambda_m = 2 \pi /
k_m = 45 \, c / \omega_{pe}\), which matches with the transverse spatial scale
of the magnetic filaments at \(t \omega_{pe} \leq 450\).

\begin{figure}[h]
\begin{center}
\includegraphics[width=8cm]{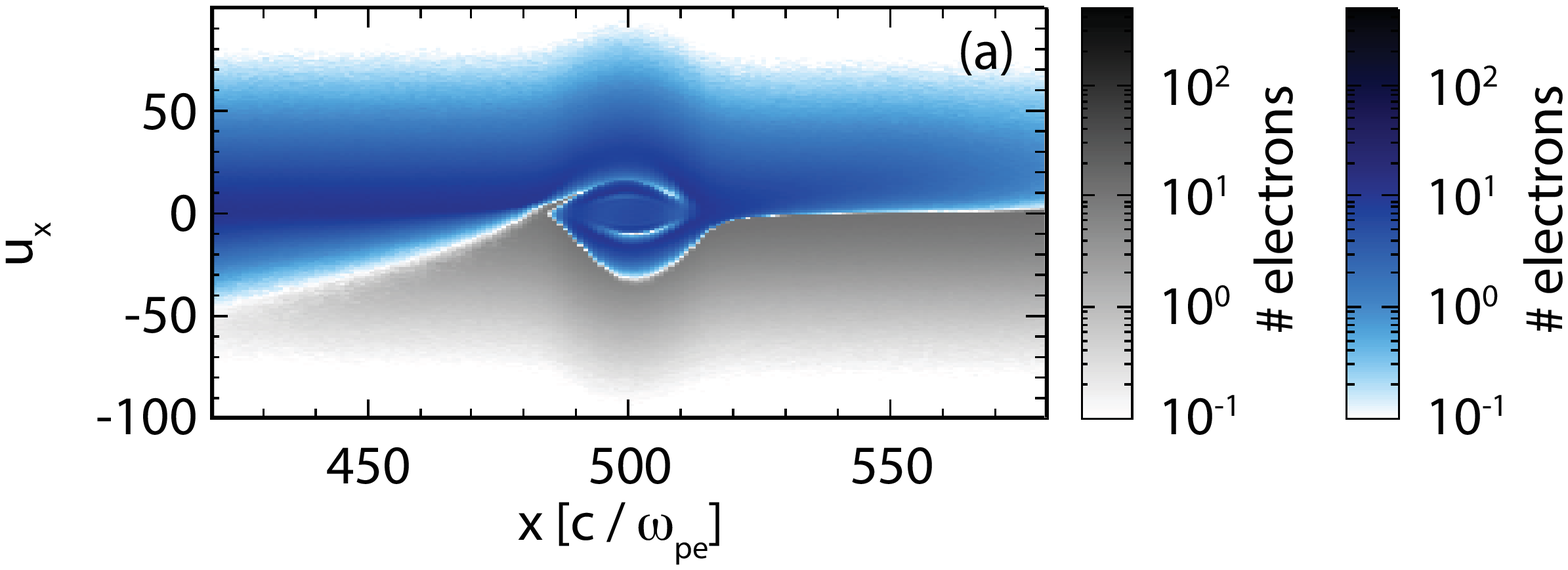}
\includegraphics[width=8cm]{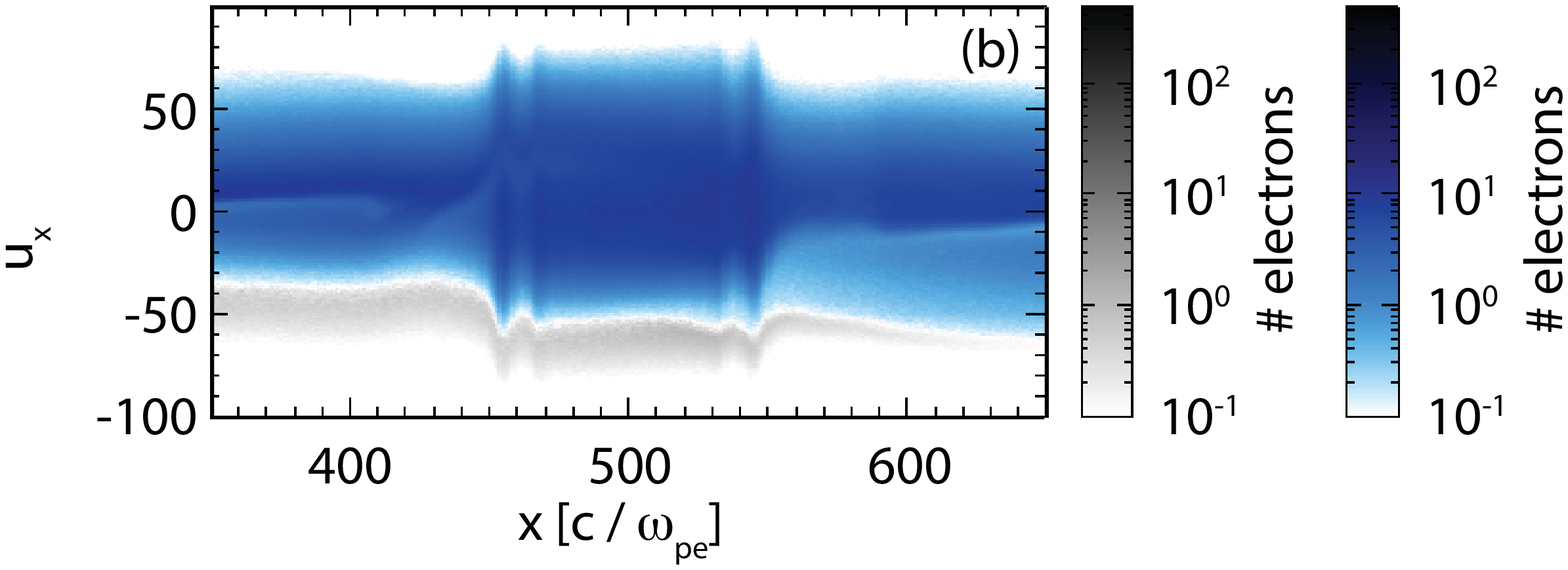}\\
\vspace{-6pt}
\includegraphics[width=8cm]{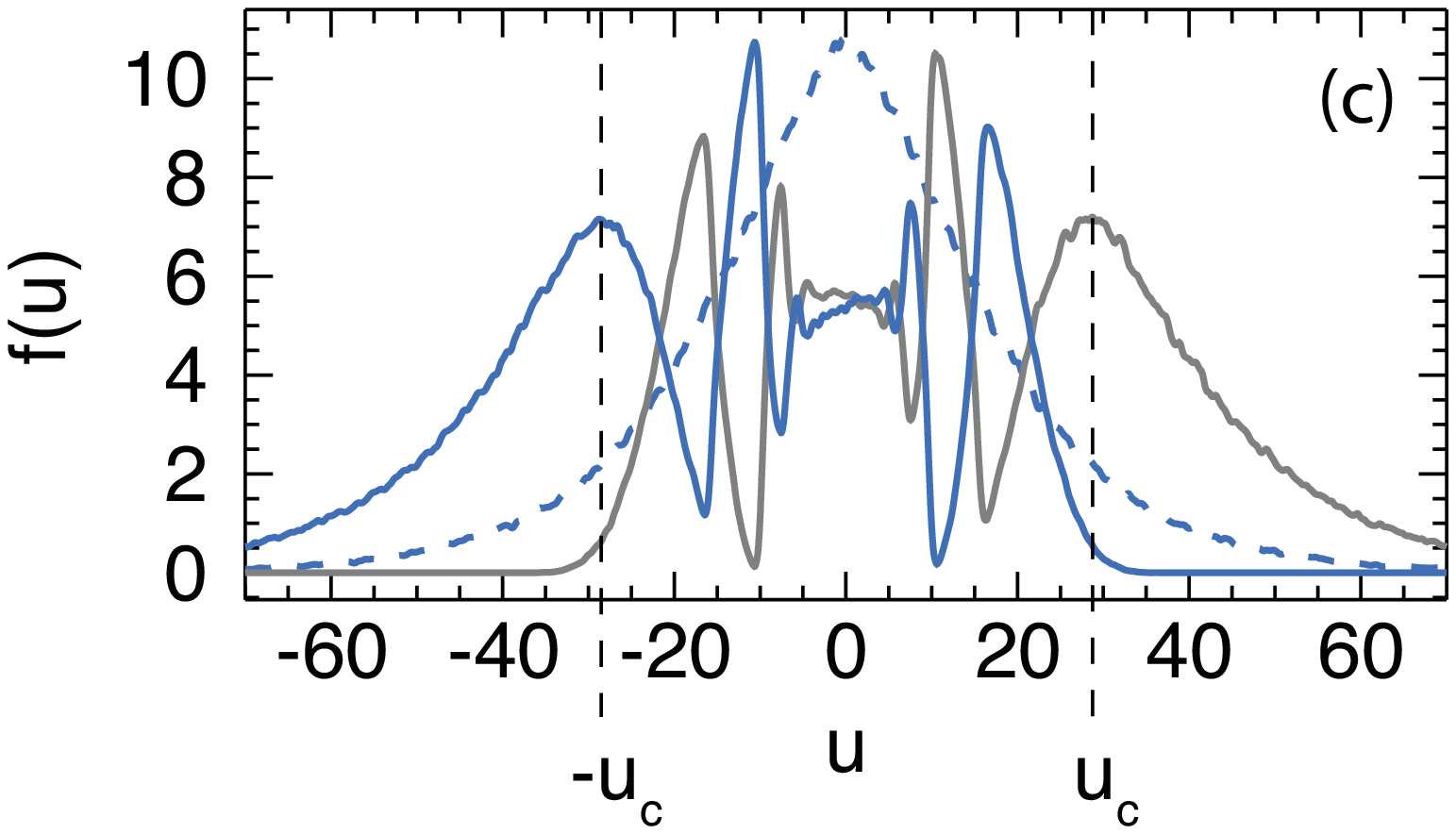}
\includegraphics[width=8cm]{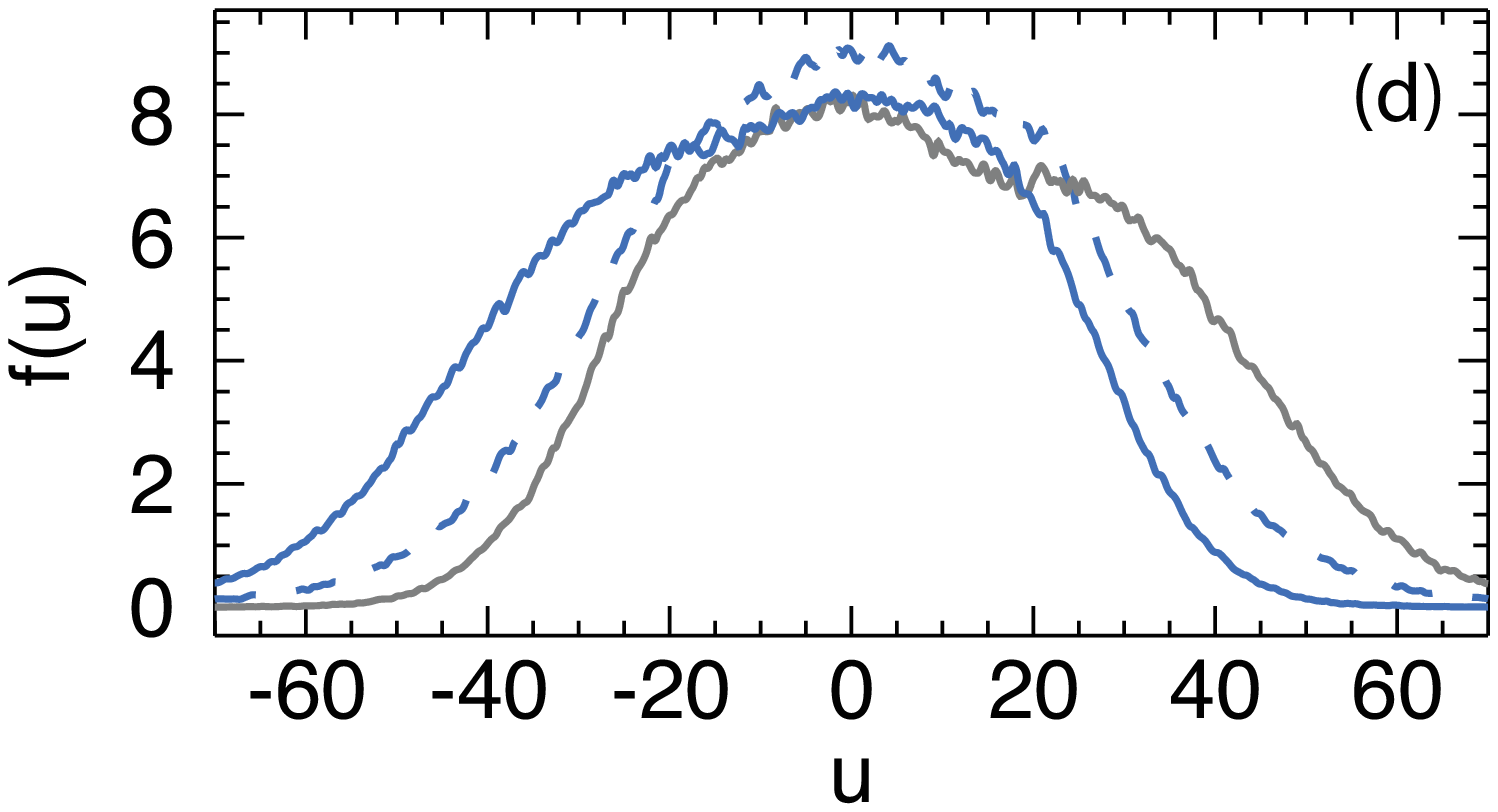}
\end{center}
\vspace{-24pt}
\caption{Electron phase spaces \((u_x, x)\) of left beam with \(u_0 = 0.1\) (blue) and right beam with \(u_0 = - 0.1\) (grey) (a, b) and electron distribution functions measured at \(x = 500 \, c / \omega_{pe}\) for momentum \(u_x\) (solid lines) for left and right beams in blue and grey (c, d), respectively,  for the 2D simulations shown in Fig.\ \ref{fig2} and for \(t \omega_{pe} = 100\) (a, c) and 1000 (b, d). The dashed lines indicate the distribution of the transverse momentum \(u_y\).}\label{fig3}
\end{figure}

The analysis of the electron distribution function explains the origin of the
anisotropy, which gives rise to the generation of the (electro)magnetic modes.
The expansion of the hot electrons relative to the slower ions creates the
strong space charge fields at the shock front. Particles are trapped in the field potential,
leading to the formation of a vortex structure in phase space as seen in figure
\ref{fig3}a, which resembles the electron holes observed in previous simulations
\cite{DS06,D08,GN08,CR12}. Along the parallel (\(\equiv\) shock propagation) direction, the two
populations of initially thermal counterstreaming beams have broadened and start
mixing, while the distribution in the perpendicular direction stays almost
unchanged (fig.\ \ref{fig3}c). This structure, with a flat distribution profile around \(u_1 = 0\) and peaks at
\( \pm u_c\), remains unaltered over hundreds of \(\omega_{pe}^{-1}\) (figs.\
\ref{fig3}b, d). At \(t \approx 1000 \, \omega_{pe}^{-1}\) the electron
distribution functions are close to thermalization (fig.\ \ref{fig2}b,
\ref{fig3}d) and the magnetic field energy has reached its quasi-steady state
(fig.\ \ref{fig2}a).

\begin{figure}[h]
\begin{center}
\includegraphics[width=7cm]{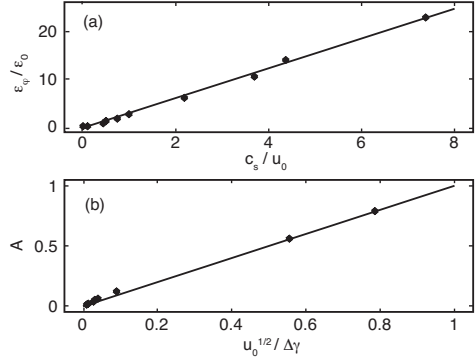}
\end{center}
\vspace{-24pt}
\caption{Parameter scan of the (a) maximum energy of the electrostatic potential normalised to the upstream
proton kinetic energy obtained from the simulations (the solid line represents a linear fit with \(c_s/u_0\)), and (b) associated maximum anisotropy \(A\) (with linear fit \(\sqrt{u_0}/\Delta \gamma\)) for 2D simulations with \(m_p/m_e = 1836\).}\label{fig4}
\end{figure}

To understand the dependence of the magnetic field generation on the properties
of the flow, we have explored the dependence of the distribution function
anisotropy \(A\) after shock formation on the flow parameters \(\Delta \gamma\) and \(u_0\). Figure
\ref{fig4} shows the scaling of the electrostatic potential and the anisotropy
with the upstream plasma parameters \(\Delta \gamma\) (or \(c_s = \sqrt{\Delta
\gamma m_e/m_p}\)) and \(u_0 \approx \sqrt{ 2 (\gamma_0 -1 )}\) in the
non-relativistic limit. Here, the parameters have been extended beyond the
electrostatic shock formation condition \(v_0 / c_s \lesssim 3\), for which the
electrostatic potential is not strong enough to form a steady-state
electrostatic shock.

It can be observed from Figure \ref{fig4}a that the electrostatic potential increases with the electron temperature and with the bulk velocity of
the upstream, \(e \varphi \propto c_s / u_0 \propto \sqrt{\Delta \gamma (\gamma_0 -1)}\)
\cite{SF13a}. Since the ion kinetic energy is a function of the initial Lorentz factor (\(\propto (\gamma_0 -1)\)), the normalisation with the initial ion energy provides a scaling \(e\varphi/\epsilon_0 \propto
\sqrt{\Delta \gamma / (\gamma_0 -1)}\), meaning that the ability of ion reflection from the electrostatic potential decreases with the initial upstream velocity. The anisotropy in figure \ref{fig4}b
shows an opposite trend; it increases with the fluid velocity and decreases with
the electron temperature with a linear dependence on
\(\sqrt{u_0}/\Delta\gamma\). Although the electrostatic potential increases
drastically with the electron temperature in the case of electrostatic shocks,
particle trapping is apparently less efficient.


The above simulations were performed for the scenario when the two beams are
initially in contact and penetrate as soon as the simulation starts. We also
considered the case where the flows are separated by a vacuum region of \(200 \, c/\omega_{pe}\) to
model the collision of two initially separated flows, as occurring in
experimental setups, taking into account the plasma expansion into vacuum. In
this case, a shock is also formed due to an electrostatic field between the hot
electrons and cold ions, and the generation of a magnetic field in the
downstream is observed with the same driving mechanism. 
In the case of beams in contact the anisotropy is higher; the amplification level after saturation is the same, but the separation leads to a decrease of the magnetic field growth rate by a factor 15. 

\begin{figure}[h]
\begin{center}
\includegraphics[width=15cm]{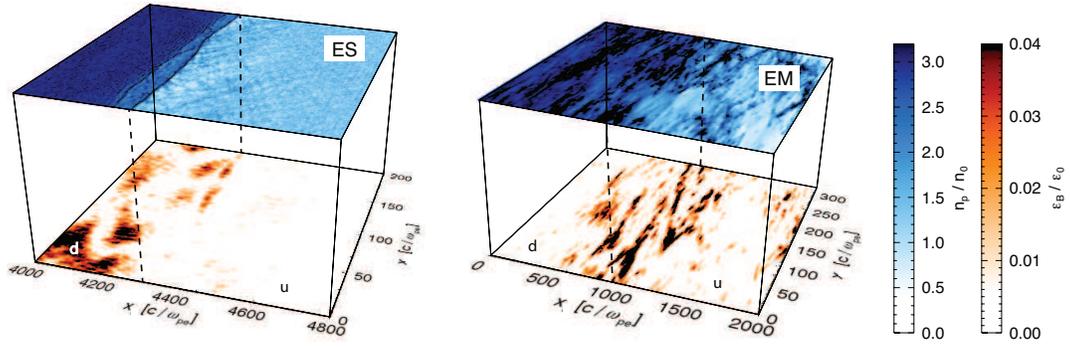}
\end{center}
\vspace{-36pt}
\caption{Proton density (blue) and normalised magnetic field energy density
(orange) obtained from 2D simulations for an electrostatic shock, ES and a Weibel-mediated shock, EM. The
dashed lines indicate the shock front with the transition from the upstream (u)
to the downstream region (d).}\label{fig6}
\end{figure}

We note that the generation mechanism for magnetic fields in the case of
electrostatic shocks is fundamentally different from the
current-filamentation-driven amplification in the case of electromagnetic
shocks. In the first case, it is actually the original Weibel instability, which
is powered by a temperature anisotropy closely tied to the shock formation process and the steady state shock structure. Therefore, the magnetic field appears in
the hotter downstream region of the electrostatic shock (see figure
\ref{fig6}). While on the other hand, in the electromagnetic case, it is the
fresh, cold upstream plasma that drives the instability and the magnetic field
appears close to the shock front, as illustrated in figure \ref{fig6} which
shows a Weibel-mediated electromagnetic shock for \(\gamma_0 = 20\) and \(\Delta
\gamma = 10^{-3}\) (parameters for both cases determined from ref.\ \cite{SF13a}).

In conclusion, we showed that strong magnetic fields are generated in the
downstream region of electrostatic shocks. Due to strong particle trapping in
the downstream region a temperature anisotropy is generated in the electron
distribution function which gives rise to electromagnetic Weibel modes. The
field clearly forms on electron time scales, and the growth rates captured in the simulations match the theoretical predictions, and with the generated field amplitude orders of magnitude
higher than it would be expected from the ion-ion filamentation instability. This
field can then seed other electromagnetic instabilities occurring on longer time
scales. We have followed the field evolution in 2D and 3D simulations and showed that
a quasi-steady state value is reached, with the magnetic field being generated only in
the downstream region, in contrary to electromagnetic shocks where the
filamentation instability creates a magnetic field across the shock front.
We observe that since the field is generated in the downstream region, the effect of the
self-generated magnetic field on the formation process is negligible, and the
properties of the electrostatic shock e.\,g.\ in terms of ion reflection are preserved \cite{HT12,FS12,FS13c}. On the other hand, the strong field
in the downstream region influences the dynamics of the particles in this region and it can lead to distinct signatures of the shock.
On the quest for the generation of collisionless shocks in the laboratory
mediated by magnetic fields, we conjecture that the identification of the structures cannot rely only on the
measurement of the self-generated magnetic fields.
Experiments will have to take into
account also the dynamics of the magnetic field generated by this mechanism in electrostatic shocks. Thus,
identification of electromagnetic shocks, in opposition to electrostatic shocks, should
rely instead on a different approach based on the direct measurement
of the relative importance of the longitudinal electric field in comparison with the the magnetic fields along the shock front.

This work was partially supported by the European Research Council (ERC-2010-AdG
Grant 267841), FCT (Portugal) grants SFRH/BPD/65008/2009, SFRH/BD/38952/2007,
and PTDC/FIS/111720/2009. The authors gratefully acknowledge PRACE for providing access to SuperMUC based in Germany at the Leibniz research center. We also acknowledge the Gauss Centre for Supercomputing (GCS) for providing computing time through the John von Neumann Institute for Computing (NIC) on the GCS share of the supercomputer JUQUEEN at J\"ulich Supercomputing Centre (JSC).


\newpage

\begin{thebibliography}{99}
\bibitem{fnote} When the shock is formed, the upstream ions are reflected by the electric field in the shock front.
\end{thebibliography}
\end{document}